\documentclass[aps,twocolumn,prb,showpacs,10pt,floatfix,groupedaddress]{revtex4-1}
\usepackage{amssymb}
\usepackage{amsmath}
\usepackage{graphicx}
\usepackage{dcolumn}
\usepackage{bm}
\usepackage{textcomp}
\usepackage{color}

\begin{document}

\title{Understanding electron behavior in strained graphene\\
as a reciprocal space distortion}

\author{M. Oliva-Leyva}
\email{moliva@fisica.unam.mx}
\author{Gerardo G. Naumis}
\email{naumis@fisica.unam.mx}

\affiliation{Departamento de F\'{i}sica-Qu\'{i}mica, Instituto de
F\'{i}sica, Universidad Nacional Aut\'{o}noma de M\'{e}xico (UNAM),
Apartado Postal 20-364, 01000 M\'{e}xico, Distrito Federal,
M\'{e}xico}


\begin{abstract}

The behavior of electrons in strained graphene is usually described
using effective pseudomagnetic fields in a Dirac equation. Here we
consider the particular case of a spatially constant strain. Our
results indicate that lattice corrections are easily understood
using a strained reciprocal space, in which the whole energy
dispersion is simply shifted and deformed. This leads to a
directional dependent Fermi velocity without producing
pseudomagnetic fields. The corrections due to atomic wavefunction
overlap changes tend to compensate such effects. Also, the
analytical expressions for the shift of the Dirac points, which do
not coincide with the $K$ points of the renormalized reciprocal
lattice, as well as the corresponding Dirac equation are found. In
view of the former results, we discuss the range of applicability of
the usual approach of considering pseudomagnetic fields in a Dirac
equation derived from the old Dirac points of the unstrained lattice
or around the $K$ points of the renormalized reciprocal lattice.
Such considerations are important if a comparison is desired with
experiments or numerical simulations.

\end{abstract}


\maketitle

\section{Introduction}

Since the experimental observation of graphene,\cite{Novoselov04} a
two-dimensional form of carbon, there have been many theoretical and
experimental studies to understand and take advantage of its
surprising properties.\cite{Geim09,Novoselov11,Neto09,Sarma} Among
its most interesting features, one can cite the peculiar interplay
between its electronic and its mechanical properties. Graphene can
withstand elastic deformations up to $20$\%, much more than in any
other crystal.\cite{Lee08} Needless to say, this long interval of
elastic response results in strong changes in the electronic
structure, which offers a new direction of exploration in
electronics: strain engineering.
\cite{Pereira09a,Pereira09b,Guinea12,Zhan} The prospect is to
explore mechanical deformations as a tool for controlling electrical
transport in graphene devices: a technological challenge owing to
the counterintuitive behavior of electrons as massless Dirac
fermions.\cite{Beenakker08}
\begin{figure}[h,t]
\includegraphics[width=8cm]{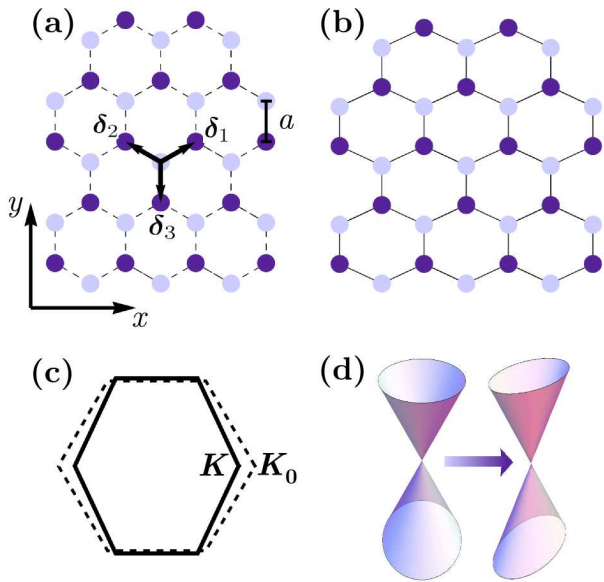}
\caption{\label{fig1} (Color online) (a) Unstrained graphene lattice
showing the vectors $\delta_i$ that point to the neighbors of type
$A$ sites, (b) the same lattice under a uniform stress, and (c), the
first Brillouin zone of the reciprocal lattice for unstrained
(dashed lines) and strained (solid lines) graphene. Note how the
reciprocal lattice is contracted in the direction where the lattice
is stretched, and the change of the $\bm{K}_0$ symmetry point  into
$\bm{K}$. (d) How the distortion  of the reciprocal lattice
transform the original Dirac cone (left) into a distorted one
(right) with a directional dependent Fermi velocity.}
\end{figure}

The most popular model proposed in the literature for studying the
concept of strain engineering is based on a combination of a
tight-binding (TB) description of the electrons and  linear
elasticity theory.\cite{Ando02,Manes07,Morpurgo,Vozmediano} In this
approach, where the absence of electron-electron interactions is
assumed, the electronic implications of lattice deformations are
captured by means of a pseudovector potential $\bm{A}$ which is
related to the strain tensor $\epsilon$ by\cite{Vozmediano}
\begin{equation}\label{VP}
A_{x}=\frac{\beta}{2a}(\epsilon_{xx}-\epsilon_{yy}), \ \ \
A_{y}=-\frac{\beta}{2a}(2\epsilon_{xy}),
\end{equation}
where $a\approx1.42$~{\AA} is the unstrained carbon-carbon distance
(see FIG.~\ref{fig1}~(a)) and $\beta\approx3$ modulates the
variation of the hopping energy $t$ of the TB model with the changes
in the intercarbon distance due to lattice
deformations.\cite{Neto09,Pereira09a} Note that, the $x$ axis is
selected parallel to the zigzag direction. This $\beta-$dependent
pseudovector potential gives a coupling of the pseudomagnetic field
($\bm{B}=\nabla\times\bm{A}$) with the electronic density. The idea
of pseudomagnetic fields has been key in the understanding of the
pseudo Landau levels experimental observations made in strained
graphene, which had been theoretically predicted
earlier.\cite{Guinea10a,Guinea10b}

In recent works, the standard description of the strain-induced
vector field has been supplemented with the explicit inclusion of
the local deformation of the lattice
vectors.\cite{Kitt,FJ,KittE,Salvador,SalvadorSSC} After accounting
for the actual atomic positions to the TB Hamiltonian, Kitt
\textit{et al.} proposed an extra pseudovector potential which is
$\beta-$independent and different at each of the strained Dirac
points.\cite{Kitt} The possible physical relevance of the extra
$\beta-$independent term, predicted in the work of Kitt \textit{et
al.}, was discussed by de Juan \textit{et al.} within the TB
approach.\cite{FJ} They also obtained an extra $\beta-$independent
pseudovector potential but with zero curl. Therefore, they concluded
that in strained graphene no $\beta-$independent pseudomagnetic
field exist.\cite{FJ,KittE}

The controversy created by Kitt \emph{et al.} has also been solved
in Refs. 21 and 22, where the concept of renormalization of the
reciprocal space was a core and novel idea developed to meet that
end. However, as has been documented in Ref. 7, the positions of the
energy minima and maxima (Dirac points) do not coincide with the
high-symmetry points at the corners of the renormalized Brillouin
zone (e.g., the $K$ point in FIG.~\ref{fig1}~(c)). This last
statement motivates us to seek the effective Hamiltonian around the
Dirac points using such renormalization, since, as shown here, this
is essential to understand the experimental data.

In this paper we analyze the most simple case, a spatially uniform
strain. The reason is that such a case must be contained as a
limiting case in any of the general theories, and at the same time,
as shown here, it can be solved exactly. Thus, it is an important
benchmark tool to compare and discriminate the goodness of previous
approaches. For example, this leads to a simple explanation for the
lattice correction terms and direction dependent Fermi velocity,
since both are due to the effects of strain in reciprocal space.
Hopefully, this will help to derive the consequences of lattice
corrections of flexural modes or curved graphene. The layout of this
work is the following. In Sec. II we present the model and find the
corresponding energy dispersion surface. In Sec. III, we discuss the
properties of the energy dispersion and find the analytical
expressions for the shift of the Dirac points and the strained Dirac
Hamiltonian. Section IV deals with the problem of how the usual
pseudomagnetic fields approach needs to be added with some
requirements in order to compare experiments and simulations.
Finally, in the last section, our conclusions are given.

\section{Model: elasticity \& tight-binding}

We are interested in uniform planar strain situations, i.e., the
components of two-dimensional strain tensor $\epsilon$ are assumed
to be position-independent. In this case, the displacement vector
$\bm{u}(\bm{x})$ is given by $\bm{u}(\bm{x})= \epsilon\cdot\bm{x}$,
and therefore, the actual position of an atom $\bm{x}'= \bm{x} +
\bm{u}(\bm{x})$ can be written  as $\bm{x}'= (I +
\epsilon)\cdot\bm{x}$, $I$ being the $2\times2$ identity matrix. In
general, if $\bm{r}$ represents a general vector in the unstrained
graphene lattice, its strained counterpart is given by the
relationship $\bm{r}'= (I + \epsilon)\cdot\bm{r}$.

We investigate the electronic implications of strain by means of the
nearest-neighbor tight-binding Hamiltonian,
\begin{equation}
H=-\sum_{\bm{x}'\!,n} t_{\bm{x}'\!,n} a_{\bm{x}'}^{\dag}
b_{\bm{x}' + \bm{\delta}_{n}'} + \text{H.c.}, \label{TBH}
\end{equation}
where $\bm{x}'$ runs over all sites of the deformed $A$ sublattice
and $\bm{\delta}_{n}'$ are the three nearest neighbor vectors. The
operators $a_{\bm{x}'}^{\dag}$ and $b_{\bm{x}' + \bm{\delta}_{n}'}$
correspond to creating and annihilating electrons on the sublattices
$A$ and $B$, at sites $\bm{x}'$ and $\bm{x}' + \bm{\delta}_{n}'$,
respectively. The dispersion relation arise upon writing
Eq.~(\ref{TBH}) in the momentum space. For this purpose, we replace
the creation/annihilation operators by their Fourier
expansions\cite{Bena}
\begin{subequations}
\begin{align}
a_{\bm{x}'}^{\dag}&=\frac{1}{\sqrt{N}}\sum_{\bm{k}_{1}}
e^{i\bm{k}_{1}\cdot(\bm{x}+\bm{u}(\bm{x}))} a_{\bm{k}_{1}}^{\dag}, \\
b_{\bm{x}'+\bm{\delta}_{n}'}&=\frac{1}{\sqrt{N}}\sum_{\bm{k}_{2}}
e^{-i\bm{k}_{2}\cdot(\bm{x}+\bm{\delta}_{n}+\bm{u}(\bm{x}+\bm{\delta}_{n}))}
b_{\bm{k}_{2}},
\end{align}
\end{subequations}
where $N$ is the number of elementary cells. In Eq.~(\ref{TBH}) we
have written the hopping integral $t_{\bm{x}'\!,n}$ as
position-dependent, but in the considered case (uniform strain), it
does not depend on the position, only on the direction:
$t_{\bm{x}'\!,n}=t_{n}$.

Under these considerations, calculation of the Hamiltonian in the
$\bm{k}$-space is fairly straightforward, $H$ becomes
\begin{equation}
H=-\sum_{\bm{k},n} t_{n}
e^{-i\bm{k}\cdot(I+\epsilon)\cdot\bm{\delta}_{n}} a_{\bm{k}}^{\dag}
b_{\bm{k}} + \text{H.c.}\label{k-H}
\end{equation}

From this equation, it follows that the dispersion relation of
graphene under spatially uniform strain is
\begin{equation}\label{GDR}
E(\bm{k})= \pm|\sum_{n}
t_{n}e^{-i\bm{k}\cdot(I+\epsilon)\cdot\bm{\delta}_{n}}|,
\end{equation}
which is a closed expression for the energy. This equation provides
a benchmark tool case for testing any Hamiltonian concerning strain
in graphene, and suggests the procedure that is developed in the
following section. If in Eq.~(\ref{GDR}), we define an auxiliary
reciprocal vector $\bm{k}^{*}= (I+\epsilon)\cdot\bm{k}$, the
dispersion relationship is almost equal to the case in unstrained
graphene, except for the different values of $t_n$ as a function of
$n$. When such hopping changes are not considered, as explained in
the following section, one gets that,
\begin{equation}\label{GDR2}
E(\bm{k})= \pm|\sum_{n} t_{0} e^{-i\bm{k}^{*}\cdot\bm{\delta}_{n}}|,
\end{equation}
which is exactly the same Hamiltonian as for unstrained graphene but
now with $\bm{k}$ replaced with  $\bm{k}^{*}$. Here, no
approximations are used and the spectrum can be obtained for all
values of $\bm{k}^{*}$. If this Hamiltonian develops around the
corresponding Dirac point, it is obvious that the same Dirac
Hamiltonian observed in unstrained graphene will appear (see below),
with $\bm{k}$ replaced with $\bm{k}^{*}$. This suggests doing a
renormalization of the reciprocal space as performed in the next
section, a result that was also found in Refs. 21 and 22.
Furthermore, Eq.~(\ref{GDR}) can be numerically evaluated to test
any effective Hamiltonian obtained by developing around particular
points in $k$ space.

It is important to remark that in the general case of a non-uniform
strain, $\bm{\delta}_{n}'$ are not given by $\bm{\delta}_{n}' =
(I+\epsilon)\cdot\bm{\delta}_{n}$. In this case, $\epsilon(\bm{x})$
needs to be replaced by the displacement gradient tensor
$\bm{\nabla}\bm{u}$.\cite{KittE,Peeters}. See in Ref. 20 how the use
of $\bm{\delta}_{n}' = (I+\bm{\nabla}\bm{u})\cdot\bm{\delta}_{n}$,
allowed  Kitt \emph{et al.} to solve the controversy concerning
whether or not lattice corrections produce pseudovector potentials.

\section{Energy spectrum of strained graphene}

The variation of the hopping energy $t_{n}$ with the changes in the
intercarbon distance fulfills a physically accurate exponential
decay $t_{n}=t_{0}\exp[-\beta(\bm{|\delta}_{n}'|/a - 1)]$, with
$t_{0}\simeq 2.7$~eV being the equilibrium hopping
energy.\cite{Neto09,Ribeiro} Nevertheless, for the sake of
comparison with previous works we consider first order in strain,
\begin{equation}
t_{n} \simeq
t_{0}(1-\frac{\beta}{a^{2}}\bm{\delta}_{n}\cdot\epsilon\cdot
\bm{\delta}_{n}).\label{aproxt}
\end{equation}

Defining the three nearest neighbor vectors as depicted in
FIG.~\ref{fig1},
\begin{equation}
\bm{\delta}_{1}=\frac{a}{2}(\sqrt{3},1), \ \
\bm{\delta}_{2}=\frac{a}{2}(\sqrt{3},1),\ \
\bm{\delta}_{3}=a(0,-1),
\end{equation}
and plugging Eq.~(\ref{aproxt}) into Eq.~(\ref{GDR}), one gets the
following expression for the dispersion relation,
\begin{equation}
E(\bm{k})=\pm t_{0}\sqrt{3 +
f(\bm{k}^*)-\beta\left(3\text{Tr}(\epsilon) +
f_{\epsilon}(\bm{k}^*)\right) +
\beta^{2} f_{\epsilon^2}(\bm{k}^*)},\label{energy}\\
\end{equation}
where $f(\bm{k}^*)$ has exactly the same functional form of its
unstrained graphene counterpart,\cite{Neto09}
\begin{equation}
f(\bm{k}^*)= 2\cos(\sqrt{3} k^{*}_{x}a) + 4\cos(\frac{\sqrt{3}
k^{*}_{x} a}{2})\cos(\frac{3 k^{*}_{y} a}{2}), \label{Eq9}
\end{equation}
but now evaluated in different points of reciprocal space, since
here $\bm{k}^{*}=({k}_{x}^{*},{k}_{y}^{*})$ is given by the transformation,
\begin{equation}
\bm{k}^{*}= (I+\epsilon)\cdot\bm{k}.
\end{equation}

This last equation is very important. It provides a mapping of the
original reciprocal space into a new distorted one. As we will see,
this mapping and the fact that $f(\bm{k}^*)$ is equal to its
undistorted counterpart lead to pure geometrical effects that only
very recently have been
identified.\cite{Kitt,FJ,KittE,Salvador,SalvadorSSC} The other terms
depend on the same distortion, but contain hopping corrections.
These terms are explicitly detailed in the Appendix~\ref{AA}.
$f_{\epsilon}(\bm{k}^*)$ contains the modification of the spectrum
due to first order in $\beta$, while $f_{\epsilon^2}(\bm{k}^*)$ is
the second order correction in $\beta$.

\subsection{Hypothetical case: $\beta=0$}

Several important consequences follow from these equations. First of
all, one can observe that in the case of deforming the lattice
without changing the hopping parameters, i.e., if one deforms the
lattice keeping $\beta=0$, $E(\bm{k})$ is simplified to:
\begin{equation}
E(\bm{k})=\pm t_{0}\sqrt{3 + f(\bm{k}^*)}. \label{purerelation}
\end{equation}

This corresponds to the same dispersion relationship observed in
graphene, but now with different reciprocal vectors, which are
obtained by applying strain to the original reciprocal vectors. In
other words, the space is strained while the eigenvalues remain the
same. As a consequence, the Dirac cone changes its shape due to this
lattice deformation, as illustrated in FIG.~\ref{fig1}. This is
exactly the result that we would obtain if a diagonalization of the
tight-binding Hamiltonian is performed using a computer. Since
$\beta=0$ and the connectivity of the lattice is not changed, the
eigenvalues of the Hamiltonian must remain equal to the undistorted
case. Only when a plot is made \emph{against the wavevectors}, the
cone turns out to be distorted, as shown in FIG.~\ref{fig1}~(d). For
example, the brick wall lattice has the same connectivity as
graphene, and thus the spectrum must be the same. However, only when
the spectrum is plotted in reciprocal space, the energy-momentum
relationship appears distorted.
\begin{figure*}[t]
{\includegraphics[width=160mm]{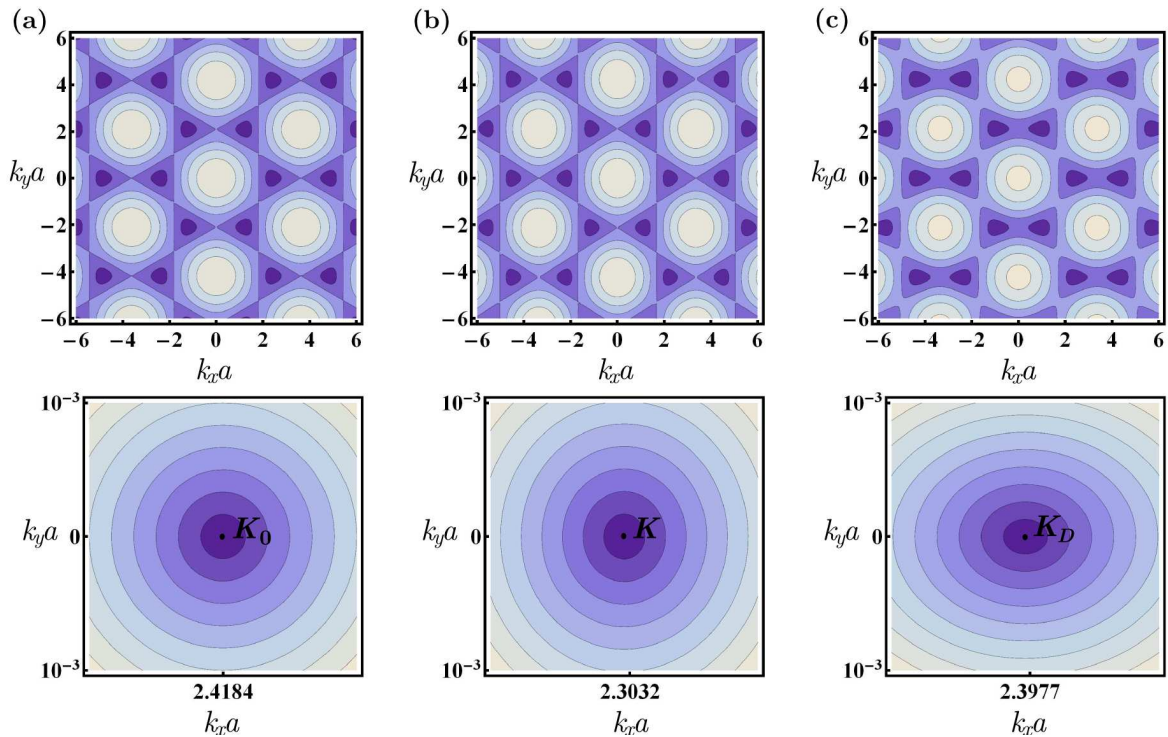}} \caption{\label{fig2} (Color
online) Isoenergetic curves obtained from the energy dispersion in
reciprocal space obtained from Eq.~(\ref{energy}). A blow up is
presented around the Dirac points $\bm{K}_D$ for each surface. Case
(a) corresponds to unstrained graphene, (b) to strained graphene
with $\beta=0, \epsilon_{xx}=0.05,
\epsilon_{xy}=0,\epsilon_{yy}=-\nu\epsilon_{xx}$, and (c) to
strained graphene with $\beta \approx 3$, and the same strain tensor
as in case (b). Note how although the strain tensor is the same in
cases (b) and (c), the ellipses are rotated by $\pi/2$, since the
reciprocal space deformation and hooping effects tend to
compensate.}
\end{figure*}

Also, the case $\beta=0$ allows us to appreciate a subtle point.
Since the spectrum is the same as in unstrained graphene, it is easy
to see that the ${\bm K}$ symmetry points of the distorted lattice
coincide with the Dirac point of the new relationship given by
Eq.~(\ref{purerelation}). In other words, the condition
$E(\bm{K}_{D})=0$, which defines the $\bm{K}_{D}$ Dirac points,
corresponds to $\bm{K}_{D}=\bm{K}$, where $\bm{K}$ is the image of
the point $\bm{K}_0$ under the mapping
$\bm{K}=(I+\epsilon)^{-1}\cdot\bm{K}_0$. Thus, for $\beta=0$ it
makes sense to develop the TB Hamiltonian around the original Dirac
points, as $\bm{k}=\bm{K}_{D}+\bm{q}$, with
$|\bm{q}|\ll|\bm{K}_{D}|$. It is easy to show that the pure
geometrical distortion allows us to write the Dirac Hamiltonian as
(see Appendix~\ref{AB}),
\begin{equation}\label{H}
H=v_0\bm{\sigma}\cdot(I+\epsilon)\cdot\bm{q}=
v_{0}\bm{\sigma}'\cdot\bm{q},
\end{equation}
$\bm{q}$ being the momentum measured relatively to the Dirac points,
$v_{0}=3t_{0}a/2$ the Fermi velocity for the undeformed lattice,
$\bm{\sigma}=(\sigma_{x},\sigma_{y})$ the two Pauli matrices and
$\bm{\sigma}'= (I+\epsilon)\cdot\bm{\sigma}$ turns out to be the
spinorial connection.\cite{Naumis} From this equation follows a
directional dependent Fermi velocity for strained graphene, which
has also been found in other works.\cite{Pereira09a,FJ} Furthermore,
a directional dependent velocity appears simply by looking at the
isoenergetic curves of Eq.~(\ref{purerelation}) around $\bm{K}$. In
this case one obtains
\begin{equation}
E(\bm{K}+\bm{q})^2=(v_0\bm{\sigma}\cdot(I+\epsilon)\cdot\bm{q})^2,
\end{equation}
therefore, the isoenergetic curves around $\bm{K}$ are rotated
ellipses, as depicted in FIG.~\ref{fig2}~(b). This figure was made
for a zigzag uniaxial strain of $5$\%, i.e., $\epsilon_{xx}=0.05,
\epsilon_{xy}=0$ and $\epsilon_{yy}=-\nu\epsilon_{xx}$, $\nu$ being
the Poisson ratio, which is very low for graphene,
$\nu\sim0.1-0.15$, according to some theoretical
estimations.\cite{Farjam,Katsnelson}

At this point, one can conclude that the basic mechanism behind the
anisotropic Fermi velocity is the distortion of the reciprocal
space. Such distortion gives a simple interpretation to the
resulting geometric crystal frame terms that appears in the
covariant version of the equations.\cite{FJ} Clearly, there are not
associated pseudomagnetic fields.\cite{Pereira09a,FJ}

\subsection{Actual case: $\beta \neq 0$}

Let us now consider the case in which the space is distorted and the
hopping is changed, i.e., $\beta \neq 0$. Here, we will have two
effects. Again one has the pure geometrical distortion due to the
strain of the reciprocal space, but at the same time, there is a
change in the spectrum. This last effect is the only one observed
when a diagonalization of the Hamiltonian is performed in a computer
for a finite number of atoms.

In FIG.~\ref{fig2}, a comparison between the cases $\beta=0$ and
$\beta \neq 0$ is presented for $E(\bm{k})$. As can be seen, the
effect of $\beta \neq 0$ is to distort the $\beta=0$ case in such a
way that it tends to compensate the strain of the reciprocal space,
i.e., the ellipses are rotated by $\pi/2$ for a realistic value of
$\beta$. The physical reason for this occurrence is that a stretched
direction in real space shirks in reciprocal space, resulting in a
higher Fermi velocity, while in the same direction, the orbital
overlap decreases since the distance between atoms increases (see
FIG.~\ref{fig1}). This tends to reduce the Fermi velocity. As a
result, lattice distortion and hopping changes tend to compensate.
This fact can also be seen in the movement of the Dirac points. From
the FIG.~\ref{fig2}, one can see that the Dirac points for the case
$\beta \neq 0$ are closer to the original ones than their $\beta=0$
counterparts.

The qualitative results discussed above, and depicted in
FIG.~\ref{fig2}, can be understood by finding analytical expressions
for $\bm{K}_{D}$ and $H$. The position of $\bm{K}_{D}$  can be
obtained from the condition $E(\bm{K}_{D})=0$. Up to first order in
strain, we obtained that $\bm{K}_{D}$ is given as follows:
\begin{equation}\label{NewKD}
\bm{K}_{D}\simeq(I+\epsilon)^{-1}\cdot(\bm{K}_{0}+\xi\bm{A})\simeq\bm{K}+\xi\bm{A},
\end{equation}
with $\bm{A}$ defined by Eq.~(\ref{VP}) and $\xi$ the valley index
of $\bm{K}_{0}$.\cite{Bena} The previous equation confirms the
remark that the Dirac points for $\beta \neq 0$ do not coincide with
the $\bm{K}$ high symmetry points of the strained Brillouin zone.
The shift, which is only produced by $\beta$, is given by the
pseudovector potential, and do not depend on $\bm{K}_0$.

Furthermore, once the points $\bm{K}_{D}$ are known, it is possible
to obtain a new Dirac Hamiltonian considering the lattice correction
and orbital overlap changes. To do this, we developed
Eq.~(\ref{k-H}) around the Dirac points using Eq.~(\ref{NewKD}), and
derived that (see Appendix~\ref{AC})
\begin{equation}\label{NewH}
H=v_0\bm{\sigma}\cdot(I+\epsilon-\beta\epsilon)\cdot\bm{q},
\end{equation}
which is a general version of Eq.~(\ref{H}), since $\beta$ effects
are included. Notice that the isoenergetic curves around
$\bm{K}_{D}$ remain ellipses, as depicted in FIG.~\ref{fig2}~(c),
but with different values of the semi-axes owing to the $\beta$
corrections. The last equation clearly shows the tendency of $\beta$
to cancel the lattice corrections.

Let us make two important remarks about Eqs.~(\ref{NewKD}) and
(\ref{NewH}), which are among the main contributions of this paper.
First, these equations are a generalization of analogous expressions
to the case of graphene under uniaxial strain which were inherited
from studies on deformed carbon nanotubes.\cite{Yang} Similar
expressions were also found for a particular case of distortion
without shear. Thus, our generalization can be reduced to other
special cases for which the results are known,
\cite{Pereira09a,Yang,Pereira10} and coincides with the exact
solvable case for $\beta=0$. Such limiting cases allows us to check
in different ways the validity of the presented results. Second,
Eq.~(\ref{NewH}) can not be derived from the theory of the
strain-induced pseudomagnetic field. Namely, the effective Dirac
Hamiltonian obtained by this theory does not reduce to our
Eq.~(\ref{NewH}) for the case of uniform strain, for reasons
explained in the following section.

\section{Experimental observation of pseudomagnetic fields}

\begin{figure}[t,h]
\includegraphics[width=8cm]{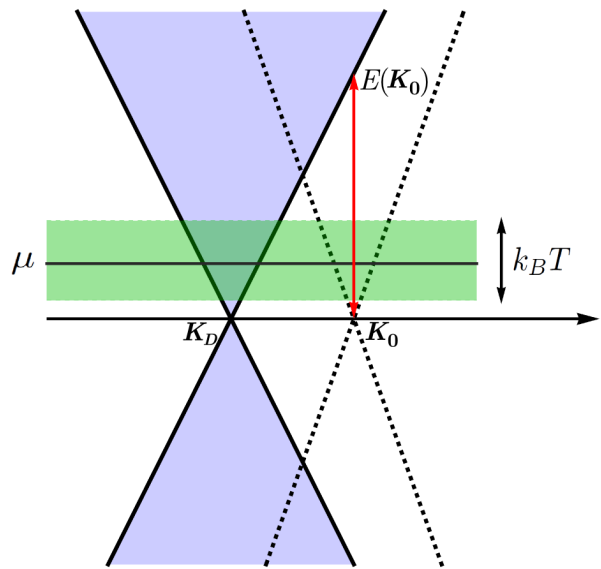}
\caption{\label{fig3} (Color online) The Dirac cone in unstrained
(dashed line) and strained graphene (solid line) and the
experimental observation of electron behavior for a probe that
shifts the chemical potential ($\mu$) with respect to the Fermi
energy. The shaded box   indicates the width of the thermal selector
due to the Fermi-Dirac distribution. The effective Dirac equation
with pseudomagnetic fields can be obtained by developing around the
original $\bm{K}_0$ points,  or in the Dirac points $\bm{K}_D$ of
the strained lattice. For $\mu=0$, only the latter approach will
work for low temperatures.}
\end{figure}

From the point of view developed in the previous section, it is
clear that basically, the Dirac cone is  translated and distorted.
As a result, if one tries to derive an effective Dirac equation
using $\bm{K}_0$ as starting points to develop $E(\bm{k})$ as
$\bm{k}=\bm{K}_{0}+\bm{q}$, the resulting energy can be quite far
away from the Fermi energy, as shown in FIG.~\ref{fig3}. This poses
a problem that has been overlooked in the usual treatment of strain
in graphene using pseudomagnetic fields in the Dirac equations.

In general, if Eq.~(\ref{energy}) is developed around a general
point in reciprocal space given by $\bm{K}_G$, we get,
\begin{equation}\label{EnergyTaylor}
E^{2}(\bm{K}_G+\bm{q})\simeq E^{2}(\bm{K}_G) +
\bm{\nabla}E^2(\bm{K}_G)\cdot\bm{q} +
\frac{1}{2}\bm{q}\cdot\bm{\nabla}\bm{\nabla}E^2(\bm{K}_G)\cdot\bm{q},
\end{equation}
where $\bm{\nabla}E^2(\bm{K}_G)$ is the Jacobian vector and $\bm{\nabla}\bm{\nabla}E^2(\bm{K}_G)$,
the Hessian matrix of $E^2(\bm{k})$, which are evaluated at $\bm{k}=\bm{K}_G$.

In the usual procedure $\bm{K}_G=\bm{K}_{0}$. However,
$E^{2}(\bm{K}_0)\neq 0$ and $\bm{\nabla}E^2(\bm{K}_G)\neq 0$. This
produces an energy shift and a $\bm{q}$ dependent term, observed in
other approaches,\cite{FJ} which complicates the description of the
dynamics somehow.

This also poses an issue concerning the experimental possibility of
observing the pseudomagnetic fields. Since the energy evaluated at
the original Dirac point $E(\bm{K}_{0})$ is different form zero, the
Fermi energy does not fall at this point, as we illustrate in
FIG.~\ref{fig3}. In general, if an experiment is performed at
temperature $T$, and the chemical potential $\mu$ is shifted by a
field, the condition to observe the pseudomagnetic fields in the
usual derivation around the original Dirac point must satisfy,
\begin{equation}
\mid E(\bm{K}_{0})-\mu \mid \leq k_BT,
\end{equation}
since the difference between $E(\bm{K}_{0})$ and $\mu$ must be less
than a zone defined from the derivative of the Fermi-Dirac
distribution against the energy, as explained in FIG.~\ref{fig3}
using a box around $\mu$. As $T \rightarrow 0$, the derivative is a
delta function centered around the Fermi energy, and the
pseudomagnetic fields calculated from $\bm{K}_{0}$ are usually far
from the region of validity. For example, even a zigzag uniaxial
strain of $1$\%, will produce a $E(\bm{K}_{0})\geq 27$~meV, which is
much higher than the thermal width of $k_BT\approx8.6$~meV, obtained
at $T=10$~K. This breaks the approximation of using pseudomagnetic
fields in a Dirac equation unless a very well defined field is used.

The option is to have a better description of the energy dispersion
near the Fermi energy, by  developing Eq.~(\ref{EnergyTaylor})
around the true Dirac points of the strained lattice, i.e., by
setting $\bm{K}_G=\bm{K}_{D}$, for which the corresponding energies
fall at the Fermi level. In this case,
\begin{equation}
E^{2}(\bm{K}_D+\bm{q})\simeq
\frac{1}{2}\bm{q}\cdot\bm{\nabla}\bm{\nabla}E^2(\bm{K}_D)\cdot\bm{q},
\end{equation}
since $E^{2}(\bm{K}_D)=0$ and $\bm{\nabla}E^2(\bm{K}_D)=0$. Now one
obtains an energy dispersion which corresponds to a distorted cone,
with a directional dependent Fermi velocity given by the elements of
the Hessian of $E^2(\bm{k})$ evaluated at $\bm{K}_D$. This result is
the same as the one obtained from the Dirac Hamiltonian given by
Eq.~(\ref{NewH}).

\section{Conclusions}

In conclusion, we have analyzed the case of a spatially uniform
strain in graphene. The lattice correction terms are simply an
effect of the strained reciprocal space. As a consequence, the Dirac
cones are deformed and translated. No pseudomagnetic fields are
associated to such terms, as has been recently
discussed.\cite{Pereira09a,FJ} When hopping changes are considered,
there is an extra deformation of the cone that tends to cancel the
effect of the reciprocal space strain. The new Dirac points of the
strained Hamiltonian do not coincide with the $\bm{K}$ symmetry
points of the strained reciprocal lattice. Due to this fact, the
effective Dirac equation can be obtained by developing around the
old or the new Dirac points. If the old points are chosen, as is
usual in the graphene literature, there is a restriction to observe
the dynamics produced by the calculated pseudomagnetic fields since
only for very high temperatures or carefully designed probes is it
possible to make a comparison with the usual theory. In computer
simulations, it is also  important to distinguish between lattice
distortion effects and connectivity matrix. Some of these issues,
can explain differences between theory and simulations in
graphene.\cite{Chang}

Finally, it is worth mentioning that although we only treated a
particular case, the ideas and lessons obtained from this study can
be translated to general cases, as we will show in forthcoming
works.

\begin{acknowledgments}
We are grateful to V.~M. Pereira and B.~B Goldberg for helpful
discussions. This work was supported by UNAM-DGAPA-PAPIIT, project
IN-$102513$. M.O.L acknowledges support from CONACYT (Mexico).
\end{acknowledgments}

\appendix

\section{}\label{AA}

In this section, we provide explicit expressions for the last terms
in Eq.~(\ref{energy}),
\begin{widetext}
\begin{align*}
f_{\epsilon}(\bm{k}^*)&= (3 \epsilon_{xx} +
\epsilon_{yy})\cos(\sqrt{3} k^{*}_{x}a) + (3 \epsilon_{xx} + 5
\epsilon_{yy})\cos(\frac{\sqrt{3} k^{*}_{x}a}{2})\cos(\frac{3
k^{*}_{y}a}{2})\notag - 2\sqrt{3} \epsilon_{xy}\sin(\frac{\sqrt{3}
k^{*}_{x}a}{2})\sin(\frac{3 k^{*}_{y}a}{2}),\\
f_{\epsilon^2}(\bm{k}^*)&=\frac{1}{8} (9 \epsilon_{xx}^2 + 6
\epsilon_{xx} \epsilon_{yy} + 9 \epsilon_{yy}^2 +
12 \epsilon_{xy}^2 + ((3 \epsilon_{xx} + \epsilon_{yy})^2 - 12 \epsilon_{xy}^2) \cos(\sqrt{3} k^{*}_{x})\\
&+ 8 \epsilon_{yy} (3 \epsilon_{xx} + \epsilon_{yy}) \cos(\sqrt{3}
k^{*}_{x}/2) \cos(3 k^{*}_{y}/2) - 16 \sqrt{3} \epsilon_{yy}
\epsilon_{xy} \sin(\sqrt{3} k^{*}_{x}/2) \sin(3 k^{*}_{y}/2)).
\end{align*}
\end{widetext}

\section{}\label{AB}

For the case $\beta=0$, the hopping integral does not depend on the
direction, $t_{n}=t_{0}$, and consequently the Hamiltonian TB given
by Eq.~(\ref{k-H}) reduces to
\begin{equation*}
H=-t_{0} \sum_{\bm{k},n}
e^{-i\bm{k}\cdot(I+\epsilon)\cdot\bm{\delta}_{n}} a_{\bm{k}}^{\dag}
b_{\bm{k}} + \text{H.c.}
\end{equation*}

The closed dispersion relation derived from this Hamiltonian has the
form
\begin{equation*}
E(\bm{k})=\pm t_{0}\sqrt{3 + f(\bm{k}^*)},
\end{equation*}
where $f(\bm{k}^*)$ is given by Eq.~(\ref{Eq9}). As  discussed in
Section II, the condition $E(\bm{K}_{D})=0$, which defines the
$\bm{K}_{D}$ Dirac points, corresponds to $\bm{K}_{D}=\bm{K}$, where
$\bm{K}$ is the image of the point $\bm{K}_0$ under the mapping
$\bm{K}=(I+\epsilon)^{-1}\cdot\bm{K}_0$. Thus, for $\beta=0$ it
makes sense to develop the TB Hamiltonian around the original Dirac
points, as $\bm{k}=\bm{K}+\bm{q}$, with $|\bm{q}|\ll|\bm{K}|$,
\begin{align*}
E(\bm{K}+\bm{q})&=\pm t_{0}\sqrt{3 + f((I+\epsilon)\cdot((I+\epsilon)^{-1}\cdot\bm{K}_{0} + \bm{q}))},\\
&=\pm t_{0}\sqrt{3 + f(\bm{K}_{0} + (I+\epsilon)\cdot\bm{q}))},\\
&=\pm t_{0}\sqrt{3 + f(\bm{K}_{0} + \bm{q}^{*}))},\ \ \ \bm{q}^{*}=(I+\epsilon)\cdot\bm{q},\\
&\simeq \pm v_{0}|\bm{q}^{*}|,\\
&\simeq \pm v_{0}|(I+\epsilon)\cdot\bm{q}|.
\end{align*}

In the next section we provide a more general proof of
Eq.~(\ref{H}). At this point, is clear that the case $\beta=0$ is a benchmark tool
for any effective Hamiltonian, since it can be solved without using any approximation.

\section{}\label{AC}

We start with the Hamiltonian in momentum space of strained graphene,
\begin{equation}
H=-\sum_{n=1}^{3}t_{n}
\begin{pmatrix}
0 & e^{-i\bm{k}\cdot(I+\epsilon)\cdot\bm{\delta}_{n}}\\
e^{i\bm{k}\cdot(I+\epsilon)\cdot\bm{\delta}_{n}} & 0
\end{pmatrix},
\end{equation}
where $t_{n}$ is given by Eq.~(\ref{aproxt}). Now, let us develop
this Hamiltonian around an original Dirac point $\bm{K}_{D}$, which
is defined by $\bm{K}_{D}=\bm{K} + \bm{A}$. Expanding
$\bm{k}=\bm{K}_{D}+\bm{q}$ we get
\begin{equation}
H=-\sum_{n=1}^{3}t_{n}
\begin{pmatrix}
0 & e^{-i(\bm{K}_{D}+\bm{q})\cdot(I+\epsilon)\cdot\bm{\delta}_{n}}\\
e^{i(\bm{K}_{D}+\bm{q})\cdot(I+\epsilon)\cdot\bm{\delta}_{n}} & 0
\end{pmatrix},
\end{equation}
but $\bm{K}_{D}\cdot(I+\epsilon)\cdot\bm{\delta}_{n}=(\bm{K}_{0} +
\bm{A})\cdot\bm{\delta}_{n}$, and to first order in $\bm{q}$ and
$\epsilon$ we may write
\begin{equation}
H\simeq-\sum_{n=1}^{3}t_{n}
\begin{pmatrix}
0 & e^{-i\bm{K}_{0}\cdot\bm{\delta}_{n}}\\
e^{i\bm{K}_{0}\cdot\bm{\delta}_{n}} & 0
\end{pmatrix}(1-i\sigma_{3}\bm{A}\cdot\bm{\delta}_{n})(1-i\sigma_{3}\bm{q}\cdot(I+\epsilon)\cdot\bm{\delta}_{n}),
\end{equation}
note that $\bm{A}$ is an expression in the first order of strain.
Using the following identity
\begin{equation}
\begin{pmatrix}
0 & e^{-i\bm{K}_{0}\cdot\bm{\delta}_{n}}\\
e^{i\bm{K}_{0}\cdot\bm{\delta}_{n}} & 0
\end{pmatrix}=i\frac{\bm{\sigma}\cdot\bm{\delta}_{n}}{a}\sigma_{3},
\end{equation}
$\bm{\sigma}=(\sigma_{x},\sigma_{y})$ being  the two Pauli matrices,
the Hamiltonian becomes
\begin{widetext}
\begin{align}
H&\simeq-t_{0}\sum_{n=1}^{3}(1-\frac{\beta}{a^2}\bm{\delta}_{n}\cdot\epsilon\cdot\bm{\delta}_{n})
(i\frac{\bm{\sigma}\cdot\bm{\delta}_{n}}{a}\sigma_{3})(1-i\sigma_{3}\bm{A}\cdot\bm{\delta}_{n}
-i\sigma_{3}\bm{q}\cdot(I+\epsilon)\cdot\bm{\delta}_{n}-(\bm{A}\cdot\bm{\delta}_{n})(\bm{q}\cdot\bm{\delta}_{n})),\notag\\
&\simeq-t_{0}\sum_{n=1}^{3}(i\frac{\bm{\sigma}\cdot\bm{\delta}_{n}}{a}\sigma_{3})(1-i\sigma_{3}\bm{q}\cdot(I+\epsilon)\cdot\bm{\delta}_{n}
-(\bm{A}\cdot\bm{\delta}_{n})(\bm{q}\cdot\bm{\delta}_{n})+
i\frac{\beta}{a^2}\sigma_{3}(\bm{\delta}_{n}\cdot\epsilon\cdot\bm{\delta}_{n})(\bm{q}\cdot\bm{\delta}_{n})
-i\sigma_{3}\bm{A}\cdot\bm{\delta}_{n}-\frac{\beta}{a^2}\bm{\delta}_{n}\cdot\epsilon\cdot\bm{\delta}_{n}),\notag\\
&\simeq v_{0}\bm{\sigma}\cdot(I+\epsilon)\bm{q} -
v_{0}\bm{\sigma}\cdot\frac{\beta}{4}(2\epsilon-\text{Tr}(\epsilon)I)\cdot\bm{q}-
v_{0}\bm{\sigma}\cdot\frac{\beta}{4}(2\epsilon+\text{Tr}(\epsilon)I)\cdot\bm{q},\notag\\
&\simeq v_{0}\bm{\sigma}\cdot(I+\epsilon-\beta\epsilon)\cdot\bm{q}.
\end{align}
\end{widetext}

This is our Eq.~(\ref{NewH}), which also reproduces Eq.~(\ref{H})
for $\beta=0$, therefore, this section can be taken as a proof of
both equations: Eq.~(\ref{H}) and Eq.~(\ref{NewH}). It is important
to emphasize that in this proof we assumed that the valley index of
$\bm{K}_{0}$ is $\xi=1$. For the case $\xi=-1$, the proof is
analogous, and the Hamiltonian is
\begin{equation}
H\simeq
v_{0}\bm{\sigma}^{*}\cdot(I+\epsilon-\beta\epsilon)\cdot\bm{q},
\end{equation}
with $\bm{\sigma}^{*}=(\sigma_x,-\sigma_y)$.

\end{document}